\documentclass[showpacs,preprintnumbers,amsmath,amssymb]{revtex4}
\usepackage{graphicx}
\usepackage{bm}
\usepackage{amsmath}

\newcommand{\be}{\begin{equation}}
\newcommand{\ee}{\end{equation}}

\begin{document}
\title{Volumetric formulation of lattice Boltzmann models with energy conservation}
\author{M. Sbragaglia$^{1}$ \& K. Sugiyama$^{2}$ \\
$^{1}$ Dipartimento di Fisica and  INFN, Universit\`a di Tor Vergata, Via della Ricerca Scientifica 1, 00133 Rome, Italy\\ 
$^{2}$ Department of Mechanical Engineering, School of Engineering, The University of Tokyo,\\ 7-3-1 Hongo Bunkyo-Ku, Tokyo 113-8656, Japan}

\begin{abstract} 
We analyze a volumetric formulation of lattice Boltzmann for compressible thermal fluid flows. The velocity set is chosen with the desired accuracy, based on the Gauss-Hermite quadrature procedure, and tested against controlled problems in bounded and unbounded fluids. The method allows the simulation of thermohydrodyamical problems without the need to preserve the exact space-filling nature of the velocity set, but still ensuring the exact conservation laws for density, momentum and energy. Issues related to boundary condition problems and improvements  based on grid refinement are also investigated.
\end{abstract}


\maketitle

\section{Introduction}

Recent studies on the lattice Boltzmann method (LBM) \cite{Chen98,BSV92} have prompted tremendous advancements in the capabilities of the method to systematically handle and reproduce complex flow properties \cite{SC93,SC94,SC95,Ansumali05,Harting08,Sbragaglia09,Wagab,Yeo95,Shan06,Chikatamarla09,Meng09}.  When dealing with isothermal Navier-Stokes equations with small degree of compressibility, LBM is frequently used with standard lattices possessing a relatively small number of velocities (less than ten in two dimensions and just a few tens in three dimensions). However, the situation is quite different and deeply challenging for compressible thermal flows  \cite{Shan06,Sbragaglia09,Watari09,Prasianakis06,Prasianakis08,Philippi06,Siebert07,Sofonea09}. As a matter of fact, it is not easy to incorporate the temperature into the lattice equilibrium when using the standard lattices, and simultaneously to satisfy a number of conditions for recovering the correct thermohydrodynamical description of compressible flows. This has triggered the development of higher order LBM schemes with larger and more isotropic sets of velocities \cite{Shan06,Philippi06,Siebert07}. Possible ways of obtaining these models are by discretizing the Boltzmann equation on the roots of Hermite polynomials and systematically derive new complete Galilean-invariant LBM schemes  \cite{Shan06,Philippi06,Siebert07,ShanHe98,HeLuo97,Martys98,Nie08},  or also introduce a systematic approach to construct higher-order lattices for stable LBM  based on the entropic approach \cite{Chikatamarla09,Chikatamarla06}. \\
Whatever is the systematic procedure used, when the roots of the velocities  are irrational, the corresponding discrete velocities cannot be fitted into a regular {\it space-filling} lattice. Thus, one of the most important advantages of the LBM, i.e. the exact space discretization of the advection step, is lost for the {\it off-lattice} models. For achieving a better accuracy, still keeping exact space-filling discretization, LBM with a large number of velocities were suggested based on the Hermite-Gauss quadrature procedure. Just to give an example, the $D2Q53$ and $D2Q81$ models detailed in a recent number of papers \cite{Siebert07,Philippi06,Philippi07}, allow for a precise higher order accuracy, but they possess much less flexibility with respect to standard models due to the increasing number of kinetic fields. Also, the use of those models with an exact discretization of the streaming step may pose the serious problem of boundary conditions, which is not an easy task when the number of velocities is increasing.\\
In order to keep a reasonably high accuracy of the lattice velocities and still retain a tractable number of them, one is somehow forced to move on off-grid lattices and find the correct computational scheme to be used. In particular, these have included interpolation schemes \cite{He96}, different finite volume schemes \cite{Peng98,Peng99a,Peng99b,Chen98} and LBM with local grid refinements and unstructured grids/adaptive meshes \cite{Filippova98,Ubertini03}. A particularly interesting approach has been discussed in a recent number of papers by Peng and coworkers \cite{Peng98,Peng99a,Peng99b}, based on finite volume techniques in the LBM framework. The resulting  lattice Boltzmann schemes  integrate the differential form of LBM using a finite-volume scheme in which the unknown populations are placed at the nodes of the mesh and evolve based on the fluxes crossing the edges of the corresponding elements. \\
In this paper we numerically and theoretically explore the potentiality of a volumetric formulation for LBM with active thermal fluctuations (hereafter refereed as TVLBM). The thermal part of the model heavily relies on Hermite quadratures with integer and non integer roots \cite{Shan06,Philippi06,Surmas09} for both bounded and unbounded flows. This kind of approach has the obvious disadvantage to loose the exact integration of the advection step as explained before. Nevertheless, one may gain in the number of used velocities that are not constrained any longer to be space-filling ones. In contrast  to point-wise interpolation schemes, this approach can be applied  without compromising exact conservation laws or equilibrium properties.  Also, due to specific properties of the methodology, the resulting TVLBM can operate on adaptive meshes, thereby providing a significant boost of geometrical flexibility especially close to the boundaries and in the properties of boundary conditions.

\section{The TVLBM model}

Our starting point is the continuum single time BGK \cite{BGK54,Chen98,Gladrow00} model written as
\be\label{eq:lb}
\partial_t f_{l}({\bm x},t)+{\bm c}_l \cdot {\bm \nabla} f_{l}({\bm x},t)=-\frac{1}{\tau} \left(f_l({\bm x},t)-f_l^{(eq)}({\bm x},t)\right)
\ee
where the left hand side represents the streaming of a probability density function, $f_l({\bm x},t)$, to find in the space-time location $({\bm x},t)$ a  particle whose velocity ${\bm v}={\bm c}_l$  is suitably chosen as belonging to a discrete set, thus enforcing the desired accuracy order. In terms of the probability density function, we can define macroscopic local variables as the density ($\rho$), velocity (${\bm u}$) or temperature ($T$)
$$
\rho=\sum_l f_{l}({\bm x},t) \hspace{.2in} \rho {\bm u}=\sum_l {\bm c}_l f_{l}({\bm x},t)  \hspace{.2in} \frac{D}{2} \rho T+\frac{1}{2} \rho u^2=\frac{1}{2} \sum_l |{\bm c}_l|^2 f_{l}({\bm x},t)
$$
with the last two equations that can be combined together to give directly the temperature
$$
\rho T=\frac{1}{D}\sum_l |{\bm c}_l-{\bm u}|^2 f_{l}({\bm x},t). 
$$
The right hand side of equation (\ref{eq:lb}) represents a single time relaxation towards a local Maxwellian equilibrium  $f^{(eq)}_{l}({\bm x},t)$ dependent on $({\bm x},t)$ via the local fields $\rho$,${\bm u}$ and $T$. In particular, for the purposes of this paper, the following  Hermite polynomials representation is adopted
$$
f_l^{(eq)}(\rho,{\bm u},T)=\omega_l \sum_{n=0}^{\infty} \frac{1}{n!}{\bm a}_{0}^{(n)}(\rho,{\bm u},T){\cal H}^{(n)}_l
$$
where
\be\begin{cases}
{\bm a}^{(0)}_0=\rho \\
{\bm a}^{(1)}_0=\rho {\bm u}  \\   
{\bm a}^{(2)}_0=\rho \left( {\bm u}^2+(T-1){\bm \delta} \right)\\
{\bm a}^{(3)}_0 =\rho \left({\bm u}^3+(T-1){\bm \delta} {\bm u}  \right)\\ 
{\bm a}^{(4)}_0=\rho \left({\bm u}^4+(T-1){\bm \delta} {\bm u}^2+(T-1)^2{\bm \delta}^2\right) \\ 
{\bm a}^{(5)}_0=\rho \left({\bm u}^5+(T-1){\bm \delta}{\bm u}^3+(T-1)^2{\bm \delta}{\bm \delta}{\bm u} \right)
\end{cases}
\ee
and the first Hermite polynomials given by
\be
\begin{cases}
{\cal H}^{(0)}_l=1 \\ 
{\cal H}^{(1)}_l={\bm c}_l \\ 
{\cal H}^{(2)}_l={\bm c}_l^2-{\bm \delta} \\ 
{\cal H}^{(3)}_l={\bm c}_l^3-{\bm c}_l {\bm \delta} \\
{\cal H}^{(4)}_l={\bm c}_l^4-{\bm c}_l^2 {\bm \delta} +{\bm \delta}{\bm \delta} \\ 
{\cal H}^{(5)}_l={\bm c}_l^5-{\bm c}_l^3 {\bm \delta} +{\bm \delta}{\bm \delta}{\bm c}_l
\end{cases}
\ee
where the shorthand notation of Grad for fully symmetric tensors has been used \cite{Shan06,Grad49}. The explicit form of the equilibrium distribution, from the standard second order in Hermite polynomials up to the fifth order, is reported in appendix A. The presence of a  single relaxation time in the evolution equation (\ref{eq:lb}) is reproducing only unitary Prandtl numbers. This pathology may be removed in different ways \cite{Shan07,Philippi07}. A simple choice may be considered the one proposed in a recent paper by Philippi and coworkers \cite{Philippi07}, where the right hand side of equation (\ref{eq:lb}) is supplemented with a local term
\be\label{FINAL}
\partial_t f_{l}({\bm x},t)+{\bm c}_l \cdot {\bm \nabla} f_{l}({\bm x},t)=-\frac{1}{\tau} \left(f_l({\bm x},t)-f_l^{(eq)}({\bm x},t)\right)+\frac{1}{\tau_g}\frac{f_l^{(eq)}({\bm x},t)}{\rho T^2} {\bm \Pi}({\bm x},t) : ({\bm c}_l -{\bm u})({\bm c}_l -{\bm u}).
\ee
The second order tensor $\Pi_{ij}$ is defined in terms of the actual fluctuations with respect to the equilibrium distributions, $f_{l}-f_l^{(eq)}$, in the following way
$$
\Pi_{ij}({\bm x},t)=\sum_l ( f_{l}({\bm x},t)-f_l^{(eq)}({\bm x},t))(c_l^i-u_i)(c_l^j-u_j).
$$
The very general scheme we will consider is therefore 
\be\label{EQEQEQ}
\partial_t f_{l}({\bm x},t)+{\bm c}_l \cdot {\bm \nabla} f_{l}({\bm x},t)=\Omega({\bm x},t)
\ee
with 
\be\label{OMEGAFVLB}
\Omega({\bm x},t)=-\frac{1}{\tau} \left(f_l({\bm x},t)-f_l^{(eq)}({\bm x},t)\right)+\frac{1}{\tau_g}\frac{f_l^{(eq)}({\bm x},t)}{\rho T^2} {\bm \Pi}({\bm x},t) : ({\bm c}_l -{\bm u})({\bm c}_l -{\bm u}).
\ee

\section{The TVLBM Evolution scheme}

\begin{figure}
\begin{center}
\includegraphics[scale=0.35]{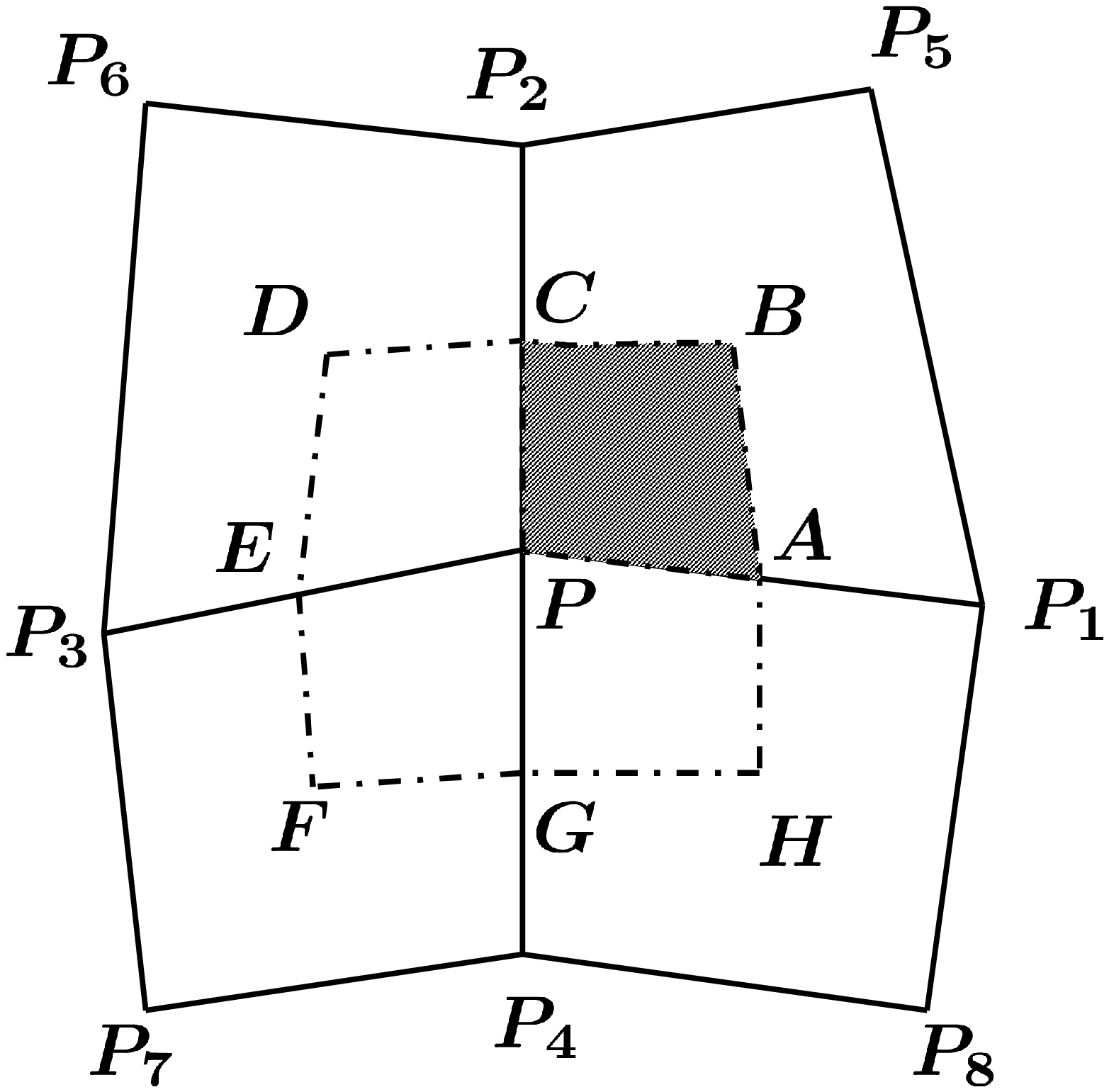}
\includegraphics[scale=0.35]{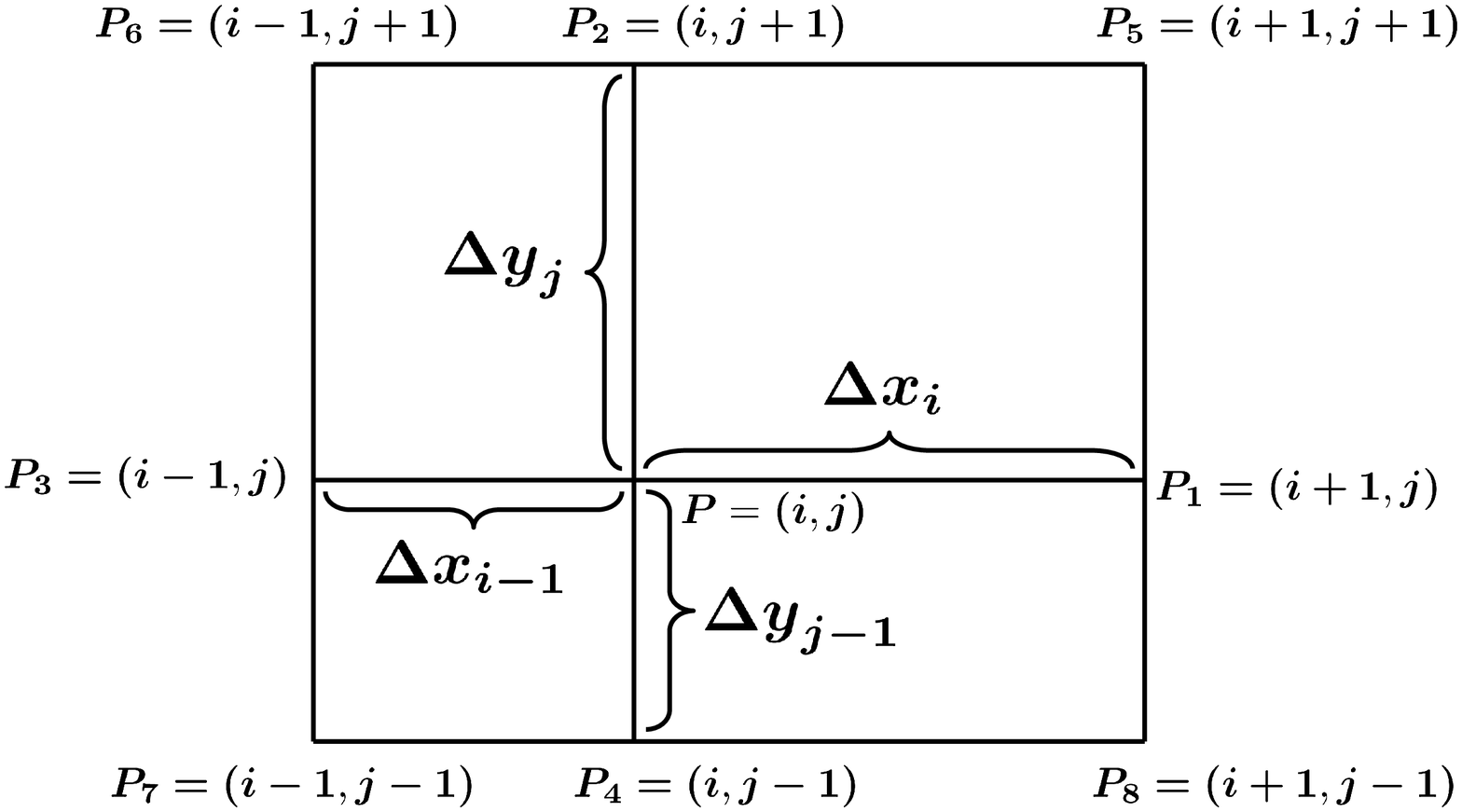}\\
\caption{Left: Diagram for finite volume implementation as reported in the papers of Peng and coworkers \cite{Peng99a,Peng99b}. The points $P$,$P_{1}$,$P_{2}$,$P_{3}$,$P_{4}$,$P_{5}$,$P_{6}$,$P_{7}$,$P_{8}$ stand for mesh points while $A$,$B$,$C$,$D$,$E$,$F$,$G$,$H$ constitute the edges  of the control volume where the integration of the lattice Boltzmann model is performed. The grey area is where the algorithm is sketched in equations (\ref{VOL1}), (\ref{VOL2}) and (\ref{VOL3}). Right: A regular (rectangular) mesh with variable spacings $\Delta x_i$ ($i=1...N_x$) and $\Delta y_j$ ($j=1...N_y$). \label{fig:scheme}}
\end{center}
\end{figure}

Following Peng and coworkers \cite{Peng99a,Peng99b}, the control volume is chosen as a generic polygon $ABCDEFGH$ (see figure \ref{fig:scheme}) surrounding the desired mesh node $P$ so that $A$, $C$, $E$ and $G$ are midpoints of edges $PP_1$, $PP_2$, $PP_3$ and $PP_4$ respectively, while $B$, $D$, $F$ and $H$ are the geometric centers of elements $PP_{1}P_{5}P_{2}$,$PP_{2}P_{6}P_{3}$,$PP_{3}P_{7}P_{4}$ and $PP_{1}P_{8}P_{4}$ respectively. We then treat the polygon $ABCDEFGH$ as made of four elements $PABC$, $PCDF$, $PEFG$, $PAHG$ and we focus on the element $PABC$ with the other integrations  done in a similar way. The various terms of equation (\ref{EQEQEQ}) are then integrated as
\be\label{VOL1}
\int_{PABC} \frac{\partial f_l}{\partial t} d \sigma= \frac{\partial f_l}{\partial t}(P) S_{PABC}
\ee
\be\label{VOL2}
\begin{split}
\int_{PABC} {\bm c}_l \cdot {\bm \nabla} f_l d \sigma=& {\bm c}_l \cdot {\bm n}_{AB} l_{AB}\frac{(f_l(A)+f_l(B))}{2}+{\bm c}_l \cdot {\bm n}_{BC} l_{BC}\frac{(f_l(B)+f_l(C))}{2}
\\&+{\bm c}_l \cdot {\bm n}_{CP} l_{CP}\frac{(f_l(C)+f_l(P))}{2}+{\bm c}_l \cdot {\bm n}_{PA} l_{PA}\frac{(f_l(P)+f_l(A))}{2}
\end{split}
\ee
\be\label{VOL3}
\int_{PABC}\Omega({\bm x},t) d \sigma=S_{PABC}\frac{(\Omega(P)+\Omega(A)+\Omega(B)+\Omega(C))}{4}.
\ee
In the above, ${\bm n}_{AB}$, ${\bm n}_{BC}$, ${\bm n}_{CP}$ and ${\bm n}_{PA}$ are the unit vectors normal to the edges $AB$, $BC$, $CP$ and $PA$, and $l_{AB}$, $l_{BC}$, $l_{CP}$ and $l_{PA}$ are the lengths of $AB$, $BC$, $CP$ and $PA$ respectively. Finally $S_{PABC}$ represents the surface area of the element $PABC$.\\
It is evident that some of the fluxes over the edges simplify and one can write down explicitly the whole evolution involving some weighted combination of the function $f_l({\bm x},t)$ in the mesh points surrounding $P$. For the purposes of this paper, we can write down  explicitly the evolution for a regular grid of $N_x \times N_y$ rectangular elements  with edges $(\Delta x)_i$ and $(\Delta y)_j$. If we consider the physical point ${\bm x}$ corresponding to the mesh point $P=(i,j)$, the evolution equation over a time lapse $dt$ is  
\be\label{FVLB}
f_l ({\bm x}, t +  d t ) = f_l ({\bm x}, t ) +  dt ( \Omega_l - {\bm c}_l \cdot {\bm \nabla} f_l ({\bm x}, t )),
\ee
and we use the following second order scheme to approximate the various quantities on the right hand side of equation (\ref{FVLB})
$$
c_l^{x} \partial_x f_{l} \approx \frac{2 c_l^x}{(\Delta x)_i+(\Delta x)_{i-1}}\left(\frac{3}{8}f_l(1)-\frac{3}{8} f_l (3) + \frac{1}{16} f_l (5) - \frac{1}{16} f_l (6) - \frac{1}{16} f_l (7) + \frac{1}{16} f_l (8)) 
 \right)
$$                
$$
c_l^{y} \partial_y f_{l} \approx \frac{2 c_l^y}{(\Delta y)_j+(\Delta y)_{j-1}}\left(\frac{3}{8}f_l(2)-\frac{3}{8} f_l (4) + \frac{1}{16} f_l (5) + \frac{1}{16} f_l (6) - \frac{1}{16} f_l (7) - \frac{1}{16} f_l (8)) 
 \right)
$$                
$$
\Omega_l \approx \left( \frac{9}{16} \Omega_l (0) + \frac{3}{32} \Omega_l (1) + \frac{3}{32} \Omega_l (2) + \frac{3}{32} \Omega_l (3) + \frac{3}{32} \Omega_l (4) + \frac{1}{64} \Omega_l (5) + \frac{1}{64} \Omega_l (6) + \frac{1}{64} \Omega_l (7) + \frac{1}{64} \Omega_l (8)  \right)
$$
where the notation $f_l(k)$ ($k=0,1,2,...,8$) has been used to identify the function $f_l({\bm x},t)$ evaluated in the mesh point $P_k$, which is indeed a first neighbor of $P=P_0$.

\section{The large scale limit}

The large scale limit of the previous TVLBM is identified with the compressible Navier-Stokes-Fourier equations for an ideal gas and non unitary Prandtl number. The technical procedure leading to such kind of equations is the well known Chapman-Enskog expansion \cite{Cowling} which is not detailed in this paper. As a matter of fact, once the kinetic equations have been correctly discretized, such a kind of procedure exactly follows the same steps of other calculations presented in the literature, both for fully continuum and lattice kinetic equations \cite{Sone2,Cercignanib,Siebert07,Buick00,Philippi07}. The final results are summarized in the following set of equations
\be\label{eq:H1}
\partial_t \rho+\partial_{i}(\rho u_i)=0
\ee
\be\label{eq:H2}
\rho\left( \partial_t u_i+u_j \partial_j u_i\right)=-\partial_{i}P+\partial_j [\eta (\partial_i u_j+\partial_j u_i-\frac{2}{D}\delta_{ij}(\partial_k u_k) )    ]
\ee
\be\label{eq:H3}
\rho\left( \partial_t T+u_j \partial_j T\right)=-P\partial_k u_k+\partial_j (2 \kappa \partial_j T )+\eta (\partial_i u_j)(\partial_i u_j+\partial_j u_i-\frac{2}{D}\delta_{ij}(\partial_k u_k) )
\ee
where we have defined the pressure
$$
P=\rho T
$$
and where the transport coefficients are given by 
$$
\eta=c_v \rho T \frac{\tau \tau_g}{2\tau+ \tau_g}; \hspace{.2in} \nu=\frac{\eta}{\rho}; \hspace{.2in} \kappa=c_P \rho T \tau
$$
with the specific heats at constant volume and pressure given by
$$
c_v=\frac{D}{2}; \hspace{.2in} c_P=\frac{D}{2}+1.
$$

\section{Measuring the transport coefficients: viscosity and thermal diffusivity}

\begin{figure}
\begin{center}
\includegraphics[scale=0.55]{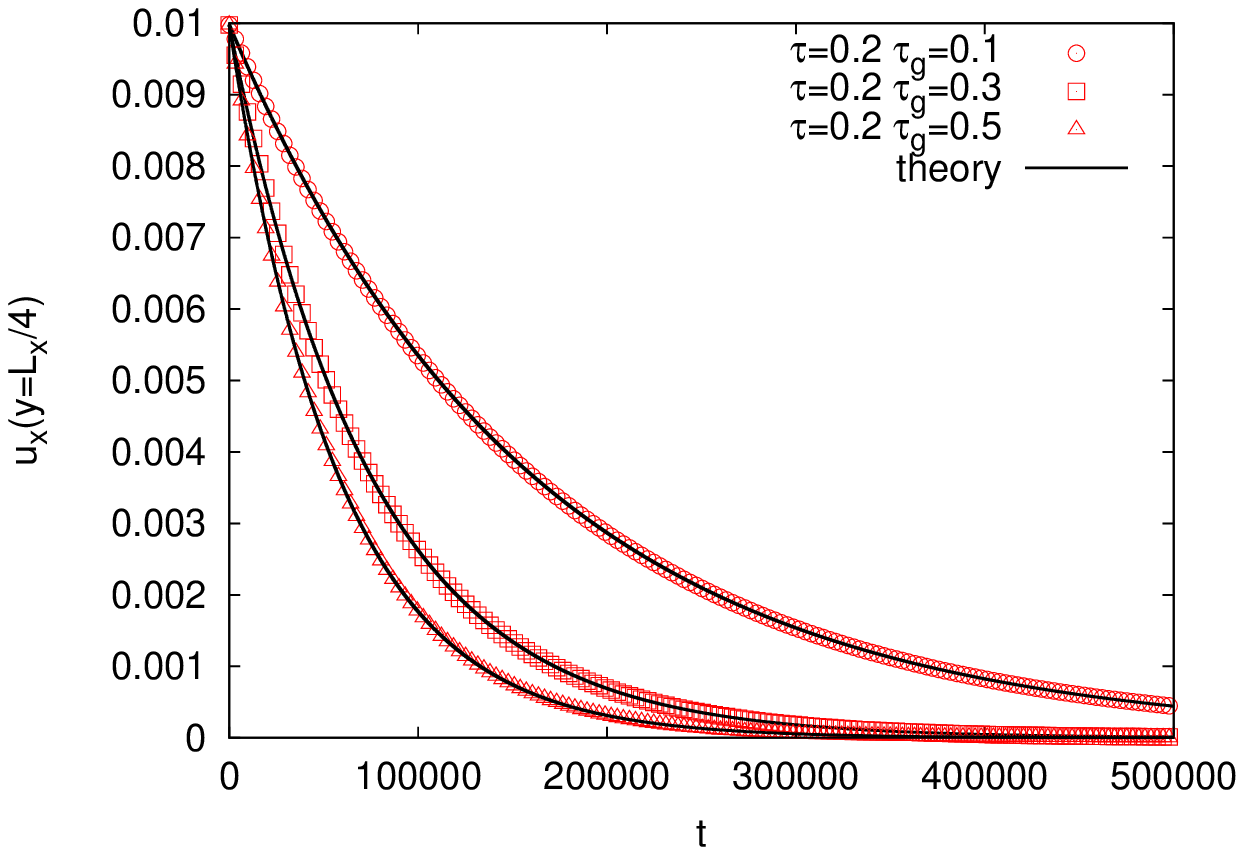}
\includegraphics[scale=0.55]{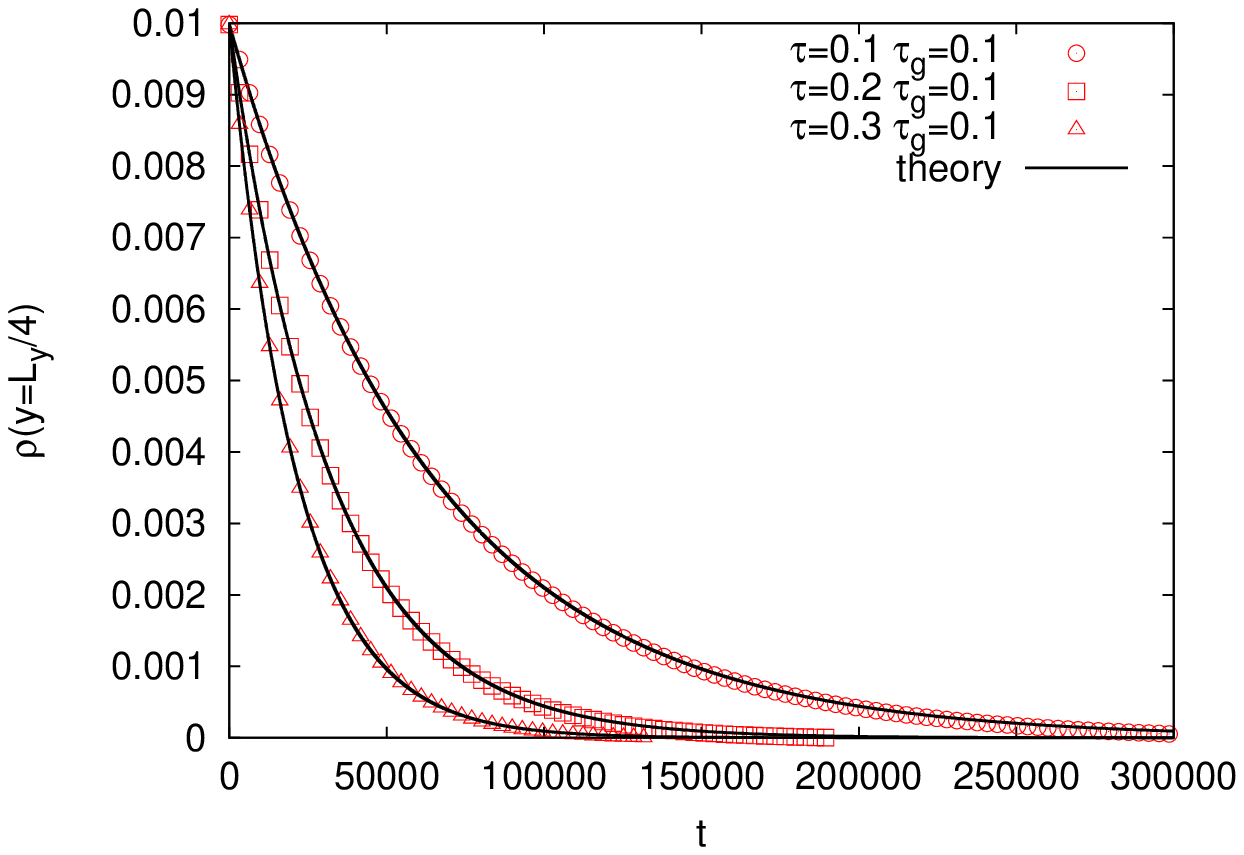}\\
\caption{Left: time history for the transverse velocity in the shear wave decay experiment in a given location $y=\frac{L_y}{4}$. Right: time history for the density in the thermal diffusion mode experiment. In both cases, the results of numerical simulations with TVLBM model given in (\ref{FINAL}) and (\ref{FVLB}) are obtained with different relaxation times as reported in the figure. The corresponding analytical prediction extracted from the linearized hydrodynamic equations (\ref{eq:H1}-\ref{eq:H3}) is also reported (solid line). The details for the initial conditions and other simulation parameters are reported in the text. \label{fig:transport}}
\end{center}
\end{figure}

The first numerical experiment we discuss is designed to measure the transport coefficients, i.e. viscosity and thermal diffusivity, thus verifying the correct convergence towards the hydrodynamical manifold (\ref{eq:H1}-\ref{eq:H3}). To avoid the complications of boundary conditions, we choose to study the evolution of the thermal diffusion and transverse shear wave modes of linearized hydrodynamics: we expect to see a wave mode decaying as $e^{-\kappa k^2 t }$ in  thermal diffusion  problems, whereas that of a transverse shear decays as $e^{-\nu k^2 t }$. All simulations are performed on a $L_x \times L_y = 0.4 \times 500$ domain, with  $N_x \times N_y = 2 \times 150$ grid points and $dt=0.005$. Three different choices of the relaxation parameters for each transport coefficient experiment are adopted: a) $\tau=0.2$ and $\tau_g=0.1$ b) $\tau=0.2$ and $\tau_g=0.2$ c) $\tau=0.2$ and $\tau_g=0.3$ for the transverse shear mode decay and  a) $\tau=0.1$ and $\tau_g=0.1$ b) $\tau=0.2$ and $\tau_g=0.1$ c) $\tau=0.3$ and $\tau_g=0.1$ for the diffusion mode decay.  For the thermal diffusion mode, the initial condition is  $\rho = \rho_0 + \epsilon \sin (2 y \pi/Ly)$, ${\bm u} = 0$, and a constant pressure. For the shear wave, the initial condition is  $u_x(x,y) = \epsilon \sin(2 \pi y/L_y)$, $u_y(x,y) = 0$, $\rho = \rho_0$. For all simulated cases, we choose $\rho_0  = 1$ and the perturbation magnitudes are set to $\epsilon = 0.01$ to ensure that we stay in the linear regime without influence of the non linear terms. The lattice velocity model used is the $D2Q21$ \cite{Surmas09} ensuring isotropy up to the eighth order tensors with $21$ off-grid velocities.  In both cases, the time histories of the perturbation magnitudes for velocity (in the kinematic viscosity measurements) and density (in the thermal diffusivity measurements) are measured and reported in figure \ref{fig:transport}. The predicted analytical behaviour is found to be well reproduced by the numerical simulations. This is a clear indication that the hydrodynamic equations are very well reproduced even with non unitary Prandtl number, as in the numerical simulations we have kept fixed one relaxation time of the model and varied the other.

\section{A test for compressibility with energy conservation: The shock tube}

\begin{figure}
\begin{center}
\includegraphics[scale=1.2]{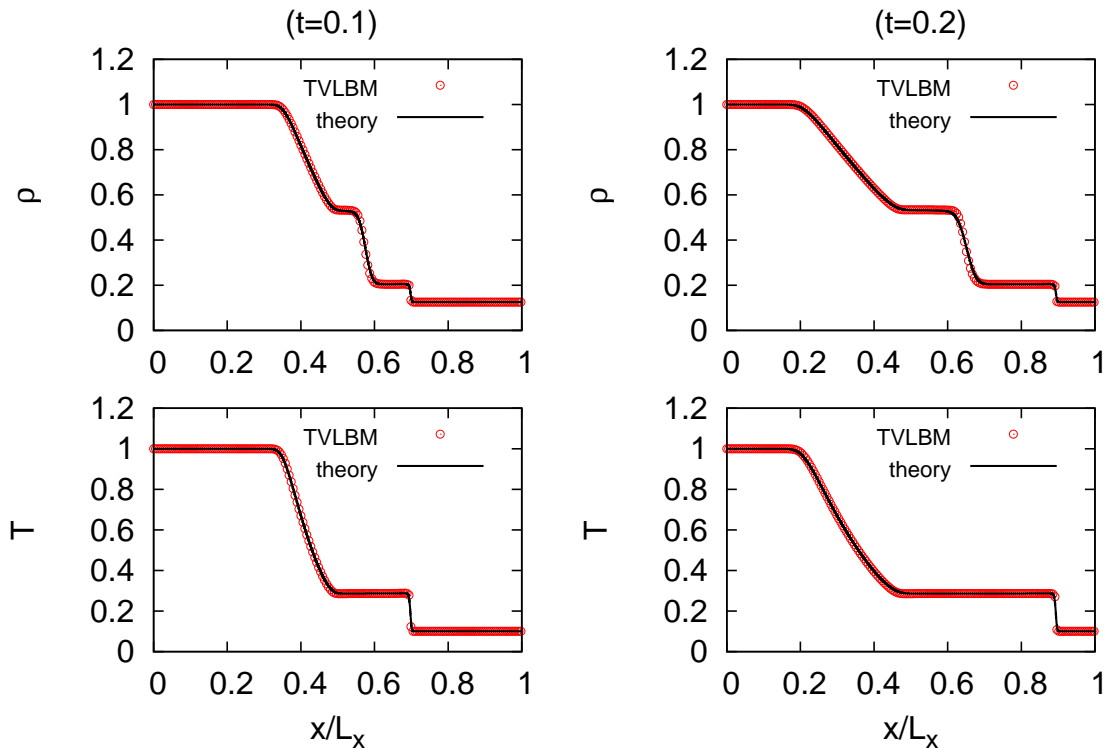}
\caption{Simulation of the Sod-Riemann shock tube. For two given times ($t=0.1$ and $t=0.2$) the numerical results obtained with TVLBM are compared with the solution of the inviscid Euler equations for a compressible gas (solid line). The numerical simulation with TVLBM is done with $21$ off-lattice velocities whose properties are reported in \cite{Surmas09}. All the other simulation details are reported in the text. \label{SOD}}
\end{center}
\end{figure}

The next numerical experiment to reveal the correct compressible thermohydrodynamical evolution  is the one-dimensional Sod-Riemann problem \cite{SOD}. We have chosen a two dimensional domain $L_x \times L_y = N_x dx \times N_y dy$ with $N_x=1500$, $N_y=2$ and $dx=dy=0.000\bar{6}$ so that $L_x=1$ and $L_y$ is so small to make the whole setup result in a one dimensional problem. Initially, the gas is at rest ($u_x=0$) with different states on the two sides of the domain's middle point: for $x \le L_x/2$ we have set $\rho=\rho_l=1.0$ and $P=P_l=1.0$ in LB units while, for $x \ge L_x/2$, we have set $\rho=\rho_r=0.125$ and $P=P_r=0.1$. This kind of initialization  is imposed in the numerics with a very sharp hyperbolic tangent profile, $\approx \tanh ((x-L_x/2)/\xi)$, separating the two half regions $x \le L_x/2$ and $x > L_x/2$, with $\xi=0.007$. In figure \ref{SOD} we compare the numerical solution with the exact theoretical prediction obtained by directly integrating the thermohydrodynamical evolution for an inviscid fluid using a finite difference Lax scheme. The solution coming from TVLBM is affected by viscous and thermal dissipation but we have chosen a very small relaxation time $\tau=0.005$  so as to ensure that, in the observed time lag, the viscous effect has negligible influence. For all the simulation we have used a single time relaxation model (\ref{FINAL}) with $\tau_g \gg 1$ and $dt=10^{-6}$. Again, the lattice velocity model used is the $D2Q21$ ensuring isotropy up to the eighth order tensors with $21$ off-grid velocities. At the edge of the segment $L_x$ we have set adiabatic boundary conditions (i.e. zero gradient) for all kinetic populations, i.e. $f_l(-dx,t)=f_l(0,t)$ and $f_l(L_x,t)=f_l(L_x+dx,t)$.

As it is clear from the figure, a rarefaction wave at left, a shock at right and a contact discontinuity at middle are observed.  In terms of constant regions and positions of discontinuity, the numerical solution agrees very well with the theoretical solution for an ideal gas, i.e. with zero viscosity and heat diffusivity. The smooth effect at the contact discontinuity and the two ends of the rarefaction wave in the numerical solution results from small but finite viscosity and heat diffusivity. When the viscosity and the heat diffusivity are reduced, the contact discontinuity will become obviously sharper.

\section{Boundary Condition: Diffuse Scattering Kernel}

In the previous sections we tested the algorithm against controlled problems of thermohydrodynamics to benchmark the correct convergence towards the Navier-Stokes-Fourier equations for unbounded fluids. The next point in order is the investigation of boundary conditions for velocity and temperature fields in the numerical simulations. We will first detail the implementation for boundary conditions inspired by the diffuse reflection concept in the rarefied theory of gases \cite{Sone1,Cercignanib,Sone2,Lorenzani,Ansumali02,Cercignania}, with benchmarks against known results for the resulting  velocity slip and temperature jump. Second, we will discuss some original ideas to implement boundary conditions in hydrodynamical problems without the emergence of velocity slip or temperature jump at the walls.\\
Let us first detail the implementation of the algorithm due to the boundaries. With reference to figure \ref{fig:scheme}, if we think the wall to be located on the node $P$, the polygons $PAHG$ and $PEFG$ have not to be included in the evolution scheme. The flux terms over the edges $PA$ and $EP$, that are basically omitted in the bulk flow implementation, are now taken into account. Given a boundary point, say ${\bm x}_W$,  the resulting scheme is given by
\be\label{FVLB2}
f_l ({\bm x}_W, t +  d t ) = f_l ({\bm x}_W, t ) +  dt ( \Omega_l - {\bm c}_l \cdot {\bm \nabla} f_l ({\bm x}_W, t )) \hspace{.3in} {\bm c }_l \cdot {\bm n} \le 0
\ee
for all those ingoing populations, i.e. those ${\bm c}_l$ such that ${\bm c }_l \cdot {\bm n} \le 0$.  As for the outgoing populations, i.e. those ${\bm c}_l$ such that ${\bm c }_l \cdot {\bm n} > 0$, we implement a boundary condition inspired by the diffuse reflection concept: the distribution functions directed to the walls mix themselves and thermalize to a local Maxwellian before getting reflected into the fluid. Before advancing the ingoing populations with (\ref{FVLB2}), we  impose the condition
\be\label{IMPOSITION}
f_l ({\bm x}_W, t)= \frac{\sum_{l, {\bm c }_l \cdot {\bm n} \le 0} |{\bm c}_l\cdot{\bm n}|  \, f_l ({\bm x}_W, t )}{\sum_{l, {\bm c }_l \cdot {\bm n} > 0} |{\bm c}_l\cdot{\bm n}| \, f_l^{(eq)}(u_w^{(eq)},T_w^{(eq)}) } f_l^{(eq)}(u_w^{(eq)},T_w^{(eq)})
\ee
for the outgoing populations. In this way, the normal velocity to the wall at time $t$ is always zero, and, when the system is propagated from $t$ to $t+dt$, we will have an inward flux of mass which exactly equals the outgoing one, i.e. the net gain of mass due to the boundaries is zero. The requirement that  condition (\ref{IMPOSITION}) is exactly satisfied every time before the advancing step implies small local depletion/gain of density. Those tiny variations, if necessary, may be balanced upon redefinition of the rest population $f_0$ . As for the details of the computational scheme, we use the following second order scheme to approximate the various quantities on the right hand side of equation (\ref{FVLB2})
$$
c_l^{x} \partial_x f_{l} \approx \frac{2 c_l^x}{(\Delta x)_i+(\Delta x)_{i-1}}\left(\frac{3}{8}f_l(1)-\frac{3}{8} f_l (3) + \frac{1}{8} f_l (5) - \frac{1}{8} f_l (6)) 
 \right)
$$                
$$
c_l^{y} \partial_y f_{l} \approx \frac{2 c_l^y}{(\Delta y)_{0}}\left(\frac{3}{8}f_l(2)-\frac{3}{8} f_l (0) + \frac{1}{16} f_l (5)+ \frac{1}{16} f_l (6)- \frac{1}{16} f_l (1)- \frac{1}{16} f_l (3) ) 
 \right)
$$                
$$
\Omega_l \approx \left( \frac{9}{16} \Omega_l (0) + \frac{3}{32} \Omega_l (1) + \frac{3}{16} \Omega_l (2) + \frac{3}{32} \Omega_l (3) + \frac{1}{32} \Omega_l (5) + \frac{1}{32} \Omega_l (6)  \right).
$$

\section{Slip and Temperature Jump for Couette Flows}\label{EXACTBC}

\begin{figure}
\begin{center}
\includegraphics[scale=0.55]{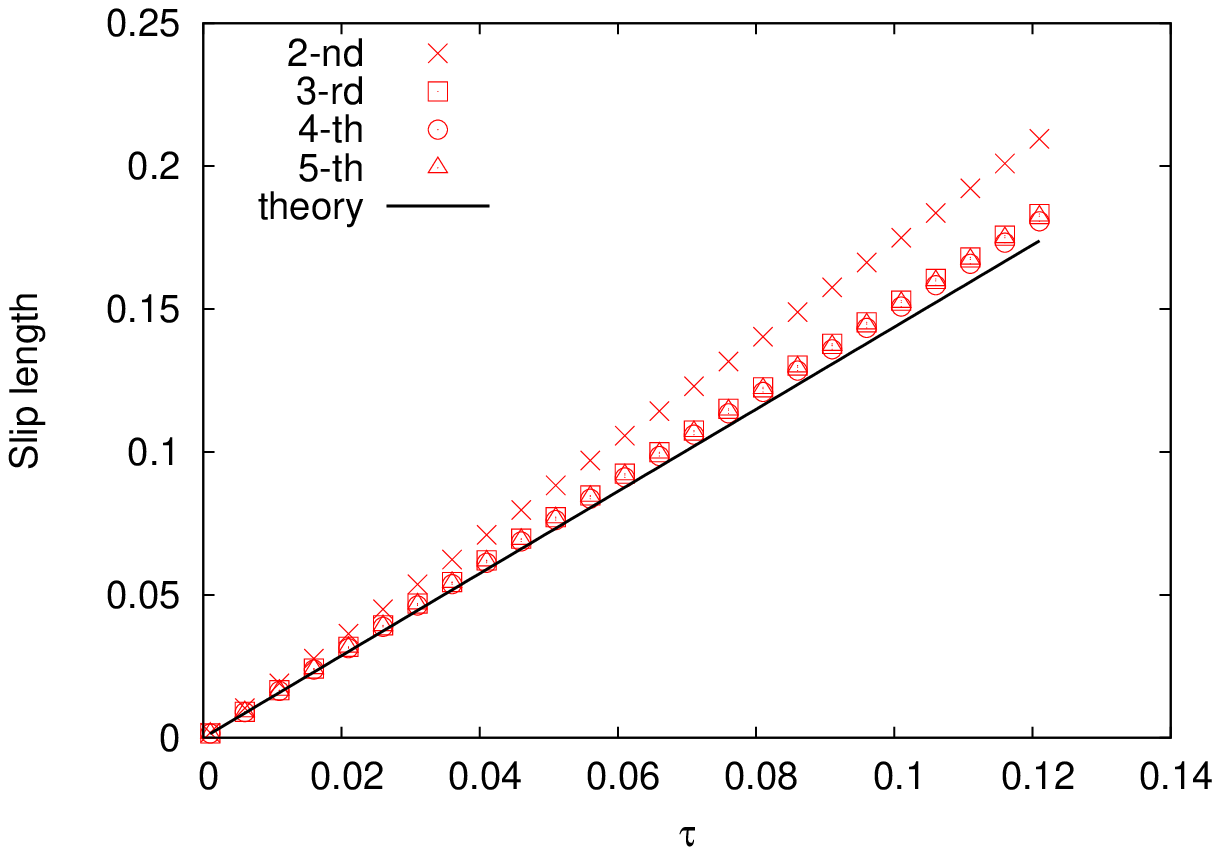}
\includegraphics[scale=0.55]{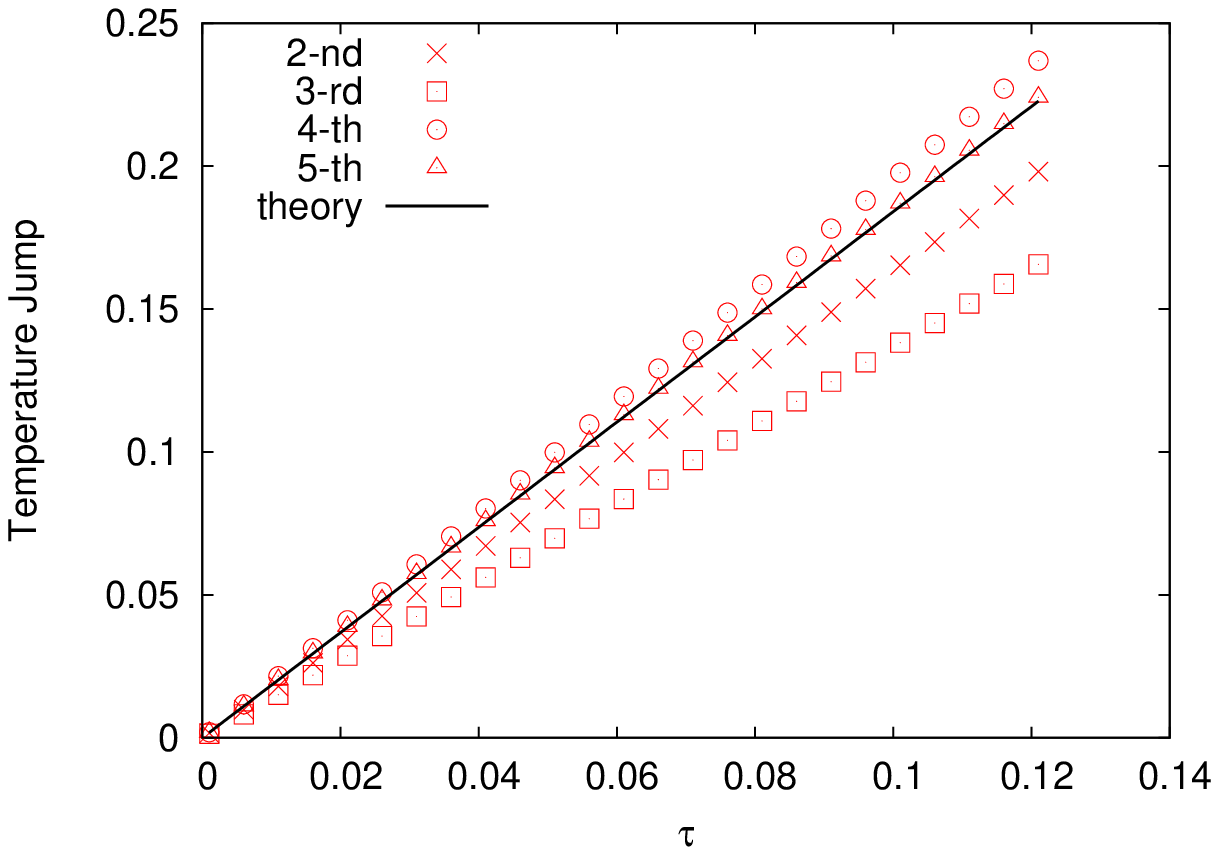}\\
\caption{Velocity slip and temperature jump from TVLBM model (\ref{FINAL}) and (\ref{FVLB}) with single time relaxation $\tau$ and diffuse boundary conditions (\ref{IMPOSITION}). Two numerical experiments are designed to compute the slip length and temperature jump emerging at the walls from the extrapolation of the profiles away from the boundary layers. Details of the numerical simulations are described in the text.  The numerical results are then compared with the result expected from a perturbative analysis of the BGK equation with such diffusive boundary condition  \cite{Sone1,Sone2} and reported in (\ref{PRED}). \label{fig:slip}}
\end{center}
\end{figure}

Given the diffuse boundary conditions (\ref{IMPOSITION}), one may want to investigate the corresponding slip velocity and temperature jump developing at the walls. For this purpose, we design two distinct experiments to test separately both effects. For the slip flow measurements we have chosen an isothermal  Couette flow with zero velocity and unitary temperature ($u_w^{(eq)}=0.0$,$T_w^{(eq)}=1.0$) in the lower wall equilibrium and a finite velocity with unitary temperature ($u_w^{(eq)}=0.01$, $T_w^{(eq)}=1.0$) in the upper wall equilibrium.  In the case of the thermal jump simulations we have set $u_w^{(eq)}=0.0$, $T_w^{(eq)}=1.005$  and  $u_w^{(eq)}=0.0$, $T_w^{(eq)}=0.995$ in the lower and upper walls respectively. We then use unitary Prandtl numbers ($\tau_g \gg 1$) with $\tau \in [0.0001:0.121]$. The simulated Couette flow has been confined in a two dimensional geometry $L_x \times L_y$ depending on the value of the relaxation parameter $\tau$. In particular, we have used $L_x \times L_y = 0.02 \frac{\tau}{\tau_0} \times 0.035 \frac{\tau}{\tau_0}$ with $\tau_0=0.001$ and the corresponding $dt$ in the simulations has been set equal to $dt=0.000001$  . The number of grid points has been kept fixed to $N_x \times N_y= 2 \times 235$. Different velocity sets (well detailed in recent papers \cite{Shan06,Surmas09}) have been used in the numerical simulations, all of them differing in  the accuracy of the Hermite polynomials of the equilibrium distribution function: a) $D2Q9$ model with nine space filling velocities  b) $D2Q12$ model with $12$ off-grid speeds and third order accuracy   c) $D2Q21$ model with $21$ off-grid speeds and fourth order accuracy   d) $D2Q28$  model with $28$ off-grid speeds and fifth order accuracy.  The corresponding results for the slip length and temperature jump are reported in figure \ref{fig:slip} and compared with the analytical prediction coming from a perturbative analysis of the BGK model \cite{Sone1,Sone2}. In particular,  the developed velocity slip and temperature jump in the aforementioned numerical experiments have been checked against the prediction
\be\label{PRED}
v_{slip}/ \left( \frac{d v}{d y} \right)= 1.43684 \, \tau  \hspace{.3in} T_{jump}/ \left( \frac{d T}{d y} \right)= 1.84074 \, \tau 
\ee
where $\left( \frac{d v}{d y} \right)$ and $\left( \frac{d T}{d y} \right)$ are the slope of the velocity and temperature profiles in the 'Navier-Stokes' region \cite{Sone1,Sone2} away from the boundary layer \footnote{This is the central region of the channel which is distant more than $\approx$ $15$ mean free paths from both walls. The theoretical prediction is extracted from equations (1.60), (3.43) and (3.47) in \cite{Sone2}}. It is evident that TVLBM correctly reproduces the desired slip velocity and temperature jump, especially in the limit of small $\tau$, where we expect the analytical prediction to work well. The importance of higher orders in the equilibrium distribution, especially to get the right temperature jump,  can be appreciated in the right panel of figure \ref{fig:slip}.

\section{Boundary Conditions: Avoiding temperature and velocity slip}

\begin{figure}
\begin{center}
\includegraphics[scale=0.75]{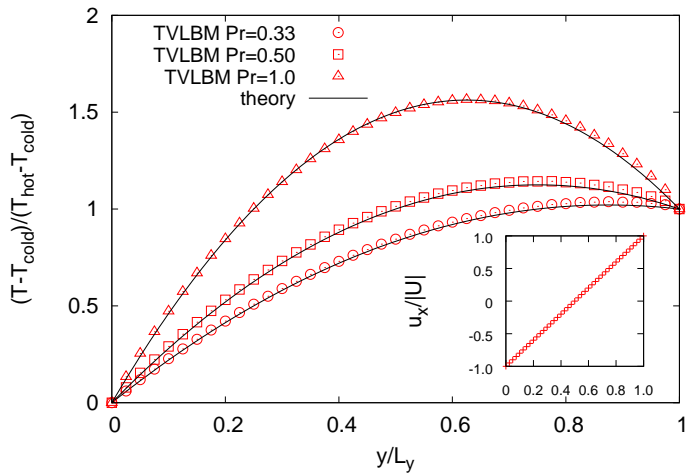}
\caption{Temperature profile for the thermal Couette flow. We have defined the normalized temperature $\frac{T-T_{cold}}{T_{hot}-T_{cold}}$, and plotted it as a function of the normalized distance from the wall $x/L_y$.  The Eckert number is kept fixed to $Ec=2.0$ while the  Prandtl number is varied between $Pr=0.3333$ and $Pr=1.0$. The corresponding analytical profiles are also shown (solid line). All the numerical results have been obtained with (\ref{FVLB}) and Dirichlet boundary conditions  imposed as explained in Section \ref{EXACTBC}. In the inset we report the shear flow in the velocity field normalized with the wall velocity. As we can see, in both temperature and velocity profiles, slip is prevented to emerge at the boundaries.  \label{ECK}}

\end{center}
\end{figure}

The previous treatment for the boundary conditions is based on the diffuse-reflection idea and, as also demonstrated before, is leading to temperature jump and velocity slip at the boundaries. It is anyhow noted that we may want to use the local parameters ($u_w,T_w$) of the local wall equilibrium $f_l^{(eq)}(u^{(eq)}_w,T^{(eq)}_w)$ to exactly impose the measured velocity and temperature at the wall, i.e. to prevent slip velocity and temperature jump. To do that, we need to impose the very same boundary condition as in the previous section with an equilibrium wall velocity and temperature chosen as $u_{w}^{(eq)}+\delta u_{w}^{(eq)}$ and $T_{w}^{(eq)}+\delta T_{w}^{(eq)}$ 
\be\label{IMPOSITION2}
f_l ({\bm x}_W, t)= \frac{\sum_{l, {\bm c }_l \cdot {\bm n} \le 0} f_l ({\bm x}_W, t )}{\sum_{l, {\bm c }_l \cdot {\bm n} > 0} f_l^{(eq)}(u^{(eq)}_w+\delta u_{w}^{(eq)},T_{w}^{(eq)}+\delta T_{w}^{(eq)}) } f_l^{(eq)}(u_{w}^{(eq)}+\delta u_{w}^{(eq)},T_{w}^{(eq)}+\delta T_{w}^{(eq)}).
\ee
The variations $\delta u_{w}^{(eq)}$ and $\delta T_{w}^{(eq)}$ are computed with an iterative Newton-Raphson procedure in such a way that the measured wall velocity and temperature are the desired ones (see Appendix B). It is computationally found that just a few iterations ($3$ or $4$) are enough to precisely set the velocity and temperature to the desired values.\\
To benchmark the new boundary conditions, the Couette flow between two infinite plates at different temperatures and velocities is simulated using the $D2Q21$ lattice with fourth order accuracy. We have chosen a two dimensional domain $L_x \times L_y = N_x dx \times N_y dy=0.2 \times 10$ with $N_x=2$ and $N_y=120$ and  the time step has been set equal to $dt=0.005$.  A relevant parameter in this case is given by the Eckert number  $Ec = U^2 /c_v \Delta T$ , where $U$ is the velocity of the upper wall, $c_v$ is the constant volume specific-heat and $\Delta T$ is the temperature difference between the walls. The velocity is set to zero in the lower wall and, accordingly with the Eckert number, different from zero in the upper wall.  The simulations were performed using a fixed Eckert number $Ec=2.0$ and variable Prandtl number $Pr$ between $0.3333$ and $1.0$. In the numerical simulations, to ensure a constant kinematic viscosity and thermal diffusivity, we have rescaled the characteristic times with the local pressure $p=\rho T$, i.e. $\tau \rightarrow \frac{\tau}{p}$ and  $\tau_g \rightarrow \frac{\tau_g}{p}$. Then, $\tau$ is kept fixed to $\tau=0.1$ and $\tau_g$ is varied according to the Prandtl number. Overall, as shown in figure \ref{ECK}, the comparison between the numerical results and the corresponding analytical estimates for the thermal Couette flows reveals that the TVLBM is able to capture correctly the expected behaviour without temperature jump and velocity slip at the boundaries.

\section{Grid Refinement for simple unidirectional flows}

\begin{figure}
\begin{center}
\includegraphics[scale=0.75]{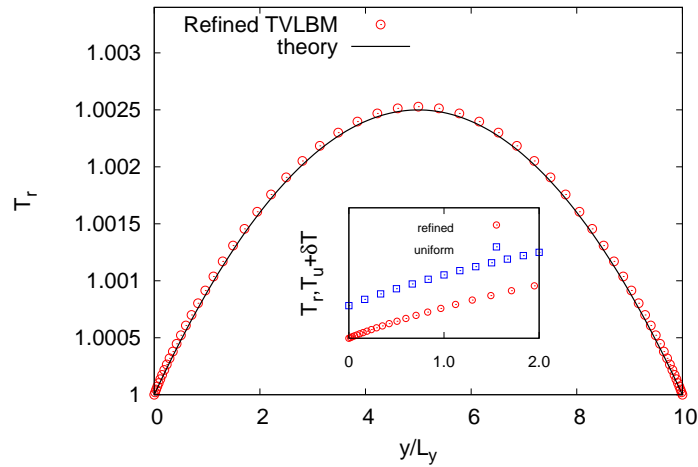}
\caption{Temperature profile for the thermal Couette flow with refined grid. The temperature is plotted as a function of the normalized distance from the lower wall. The numerical results  have been obtained with (\ref{FVLB}) and Dirichlet boundary conditions for the hydrodynamic fields, imposed as explained in Section \ref{EXACTBC}: a fixed wall velocity $U=\pm 0.1$ on both walls (located at $0$ and $y/L_y=1$) and a unitary Prandtl number have been used. The corresponding analytical profile is also shown (solid line), as predicted from stationary hydrodynamics. Also, we have used a non uniform grid normal to the walls, whose details are reported in (\ref{NONUNI1}-\ref{NONUNI2}). In the inset, we highlight the effect of grid refinement close to the lower boundary layer, and compare it with the corresponding numerical simulation with the same number of grid points arranged in a uniform way (details are reported in the text). To make it a clear distinction between the two profiles, we have shifted the profile with uniform grid ($T_u$) with respect to the one with refined grid ($T_r$) by a constant $\delta T=0.001$. \label{fig:refined}}
\end{center}
\end{figure}

In this section we explore the possibility to use TVLBM with variable grid mesh in a very simple and controlled problem involving thermal hydrodynamics. We choose a thermal Couette flow between two walls at the same temperature $T=1$ with a shear flow imposed by fixing the velocity of the upper/lower wall to $U= \pm 0.1$.  The computational setup is chosen as a two dimensional one with $L_x \times L_y= 0.2 \times 10.0$ where the streamwise length has been set equal to $L_x =N_x dx$ with $N_x=2$, $dx=0.1$ and periodic boundary conditions along it, whereas the vertical length has been covered with $N_y=60$ grid spacings satisfying
\be\label{NONUNI1}
\Delta y_{j}=y_{j+1}-y_{j} \hspace{.2in} j=1,2,3,...,N_y
\ee
\be\label{NONUNI2}
y_j=\frac{L_y}{2} \left( 1+\frac{\tanh \left( \beta \phi_j \right)}{\tanh \beta} \right) \hspace{.2in} \phi_j=\left( 1-2\frac{j-1}{N_y-1} \right)
\ee
with $\beta > 1$ a parameter determining the degree of non uniformity (see also figure \ref{fig:scheme}) of the mesh, i.e. the larger is $\beta$ the higher is the non uniformity.  For simplicity, we use a unitary Prandtl number obtained with $\tau_g \gg 1$ and $\tau=0.01$. The time step $dt$ has been chosen equal to $dt=0.005$ and the  $D2Q21$ model with $21$ off-grid speeds and fourth order accuracy has been used. As for the non uniform grid, we have chosen $\beta=2.3$ with a resulting grid spacing ranging from $dy=0.016631$ close to the boundaries up to $dy=0.391$ in the middle of the channel. Results are reported in figure \ref{fig:refined}, where the refined numerical profile for the temperature ($T_r$) is compared with the prediction coming from stationary hydrodynamics (\ref{eq:H1}-\ref{eq:H3}). The temperature profile  from this non uniform grid is also compared with the temperature profile ($T_u$) coming from a uniform grid with spacing $dy=0.1666$ at fixed $N_y$. To make it visible the effect of refinement, one profile has been shifted uniformly with respect to the other with a quantity $\delta T=0.001$. In all the numerical simulations, to ensure a constant kinematic viscosity and thermal diffusivity, we have rescaled the characteristic times with the local pressure $p=\rho T$, i.e. $\tau \rightarrow \frac{\tau}{p}$.

\section{Grid Refinement in developed RB convection}

\begin{figure}
\begin{center}
\includegraphics[scale=0.6]{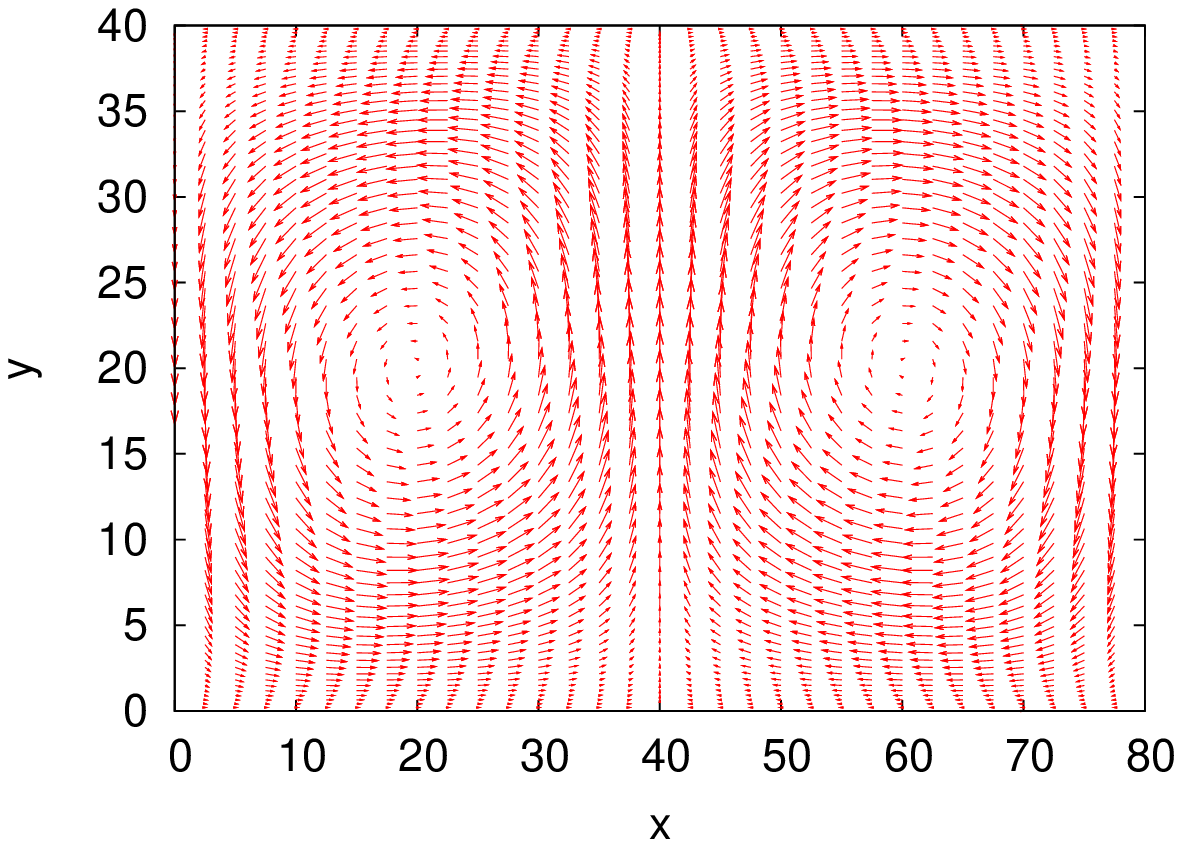}
\includegraphics[scale=0.6]{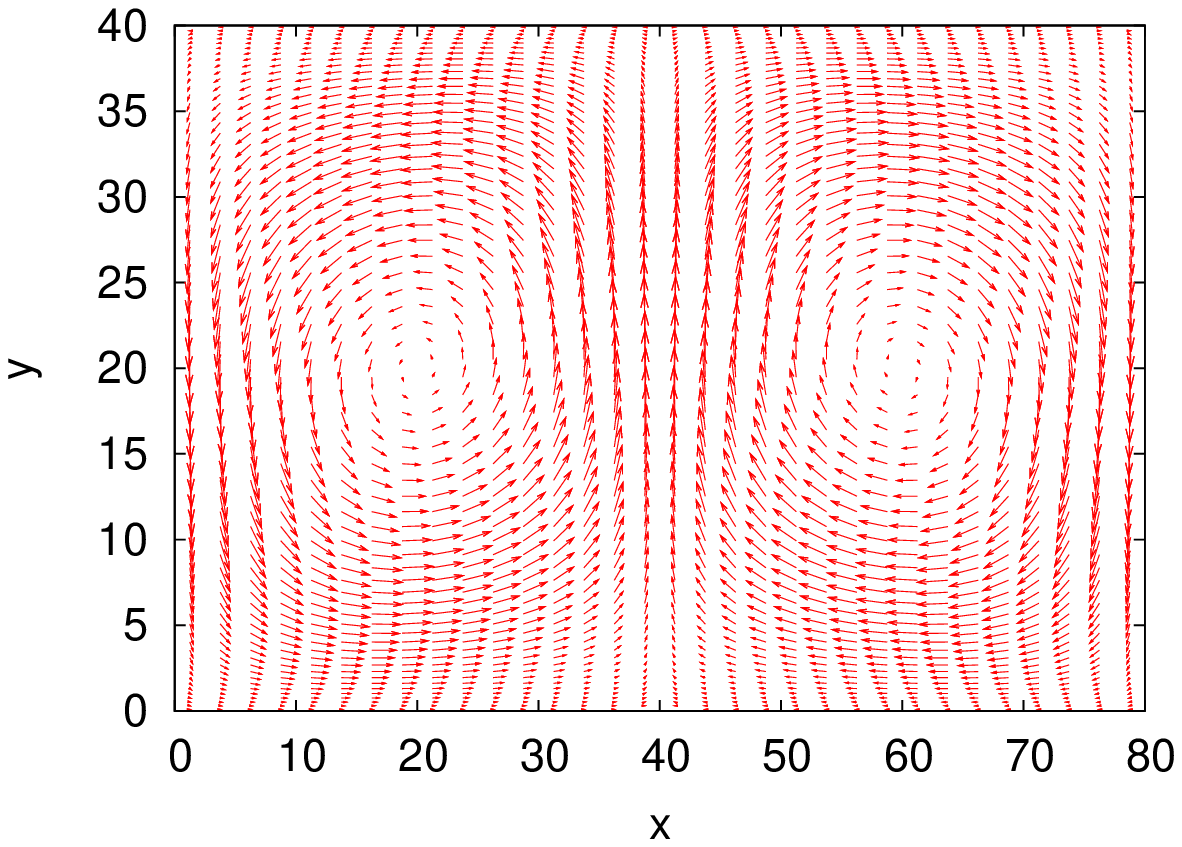}
\caption{Stationary velocity vector profiles for  Rayleigh-B\'{e}nard  thermal convection. The results from TVLBM (left) are compared with those from finite difference methods (right). The simulated system has a size $L_x \times L_y= 80 \times 40$, unitary Prandtl number $Pr=1$, and Rayleigh number $Ra=8424$. Other computational details are given in the text.
\label{fig:ROLLS}}
\end{center}
\end{figure}

\begin{figure}
\begin{center}
\includegraphics[scale=0.75]{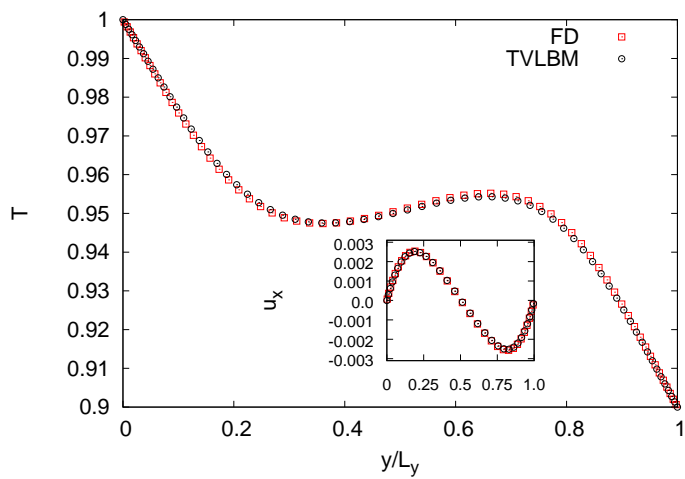}
\caption{Temperature and streamwise velocity profiles extracted from  the stationary snapshots reported in figure \ref{fig:ROLLS}. We have chosen a fixed  $x=x_0=21.25$ and we have plotted the temperature (main figure) and streamwise velocity profile (inset) from TVLBM and finite difference (FD) simulations. \label{fig:PROFILES}}
\end{center}
\end{figure}

In this section we probe the robustness of the algorithm in some non trivial two dimensional setup where thermal fluctuations are present, together with non uniform grid spacings. The setup chosen is two dimensional Rayleigh-B\'{e}nard convection \cite{Ahlers09,Siggia94,Chandrasekhar61} between two heated walls with different temperatures above the transition point, where convective rolls are present and stationary. As a matter of fact, the use of a volumetric formulation may become a valuable choice to investigate turbulent convection where we need to well resolve the boundary layer physics. The use of an exact stream and collide structure for thermal lattice Boltzmann codes may cause an error source in determining the physical properties of the boundary layer due to  the presence of spurious, small, departure from the exact linear profile in the mean temperature close to the boundary walls \cite{Scagliarini}. This departure goes together with the existence of small spurious transverse velocity  for two-three grid layers close to the wall and are due to the existence of discrete velocities which connects up to three layers in the lattice inducing non-local boundary conditions effects. Such effects can be annoying for the investigation  of highly turbulent regimes, where the boundary layer dynamics becomes crucial to drive the correct thermal exchange with the bulk \cite{verzicco}. It is numerically observed that this shortcoming can be strongly reduced by moving from LBM algorithms using exact streaming to TVLBM based on finite-volume schemes as proposed here.\\ 
In what follows, TVLBM numerical simulations are compared against results obtained using finite difference (FD) codes for the incompressible case (full details are reported in \cite{kazu1,kazu2}). The computational setup is chosen as a two dimensional box $L_x \times L_y= 80 \times 40$ where the streamwise length has been covered with $N_x=32$ points with periodic boundary conditions, while the vertical length has been covered with $N_y=64$ points and a non uniform grid with details reported in (\ref{NONUNI1}-\ref{NONUNI2}) with $\beta=1.5$ . Also, the use of a gravitational acceleration $g$ is needed for thermal convection. To do that, we implement a general forcing term in the kinetic equations with its exact representation (see for example equation (3.15) in \cite{Shan06}). In figures \ref{fig:PROFILES} and \ref{fig:ROLLS} we make a one-to-one comparison of TVLBM with FD. The TVLBM parameters (temperature difference between cold and hot walls, gravity etc...) have been set in such a way to not produce strong compressible effects with the same transport coefficients and convection intensity in both codes. In particular, the top/bottom wall temperatures have been set equal to $T_{b}=0.9$ and $T_u=1.0$, with the gravitational acceleration equal to $g=0.0001$.  The transport coefficients correspond to a unitary Prandtl number $Pr=1$ and  Rayleigh number $Ra=8224$. The stationary snapshots of the velocity vector field are reported in figure \ref{fig:PROFILES} where we see a net satisfactory agreement between the two numerical simulations. Further insight is gained by checking the details of the thermohydrodynamical profiles for a fixed $x$ as a function of $y$ in figure \ref{fig:ROLLS}. The stationary profiles are very well superposing, as shown for both temperature and velocity field in the streamwise direction.  

\section{Conclusions}

We have discussed  a volumetric formulation of lattice Boltzmann for compressible fluid flows with active thermal fluctuations (TVLBM). The model has been shown to reproduce correctly the large scale behaviour given by the Navier-Stokes-Fourier dynamics with and without boundary conditions. The velocity set has been chosen consistently with a Gauss-Hermite quadrature and is not necessarily constrained to be a space filling set, thus reducing in number the minimal set needed to obtain the correct hydrodynamic behaviour  without compromising exact conservation laws or equilibrium properties.  Also, due to specific properties of the methodology, the resulting method can easily work on adaptive  meshes, thereby providing a significant boost of geometrical flexibility.  For example, it would be extremely interesting the study of compressible thermal convection at very high Rayleigh numbers \cite{Ahlers09}, especially close to the boundaries, where the properties of the thermal boundary layer need to be well resolved to determine the input of heat into the system. At the same time, issues related to the generalization of TVLBM to non ideal gases and multiphase fluid flows have not been explored systematically in the literature, and interesting lines of research may be envisaged \cite{Brennen,Ahlers09,oresta} for the future.

\section*{Appendix A}

In this appendix we report the details for the equilibrium distribution with successive approximations, from the standard second order approximation up to the fifth order one. Given the space dimensionality $D$, the various terms entering the following definition of the equilibrium 
$$
f_l^{(eq)}(\rho,{\bm u},T)=\omega_l \sum_{n=0}^{\infty} \frac{1}{n!}{\bm a}_{0}^{(n)}(\rho,{\bm u},T){\cal H}^{(n)}_l
$$
are given by the following expressions
\be\begin{split}
\frac{1}{2!}{\bm a}_{0}^{(2)}(\rho,{\bm u},T){\cal H}^{(2)}_l =\frac{1}{2}\left(u_l^2-u^2+(T-1)(c_l^2-D) \right)
\end{split}
\ee

\be\begin{split}
\frac{1}{3!}{\bm a}_{0}^{(3)}(\rho,{\bm u},T){\cal H}^{(3)}_l =\frac{u_l}{6} \left(u_l^2-3 u^2+3 (T-1)(c_l^2-D-2) \right)
\end{split}
\ee

\be\begin{split}
\frac{1}{4!}{\bm a}_{0}^{(4)}(\rho,{\bm u},T){\cal H}^{(4)}_l =\frac{u_l^4-6 u_l^2 u^2+3 u^4}{24}+\frac{(T-1)}{4}((c_l^2-D-2)(u_l^2-u^2)-2  u_l^2)+\\
\frac{(T-1)^2}{8} (c_l^4-2(D+2) c_l^2+D(D+2))
\end{split}
\ee

\be\begin{split}
\frac{1}{5!}{\bm a}_{0}^{(5)}(\rho,{\bm u},T){\cal H}^{(5)}_l =\frac{\rho}{120} \left(u_{l}^5-10 u^2 u_{l}^3+15 u^4 u_{l} \right)+ \frac{\rho}{12} (T-1) \left(c_l^2 u_{l}^3-(D+6) u_{l}^3-3 c_l^2 u_{l} u^2 +(3 D+12) u^2 u_{l} \right)+\\
\frac{\rho}{8} (T-1)^2 \left(u_{l} c_l^4-(2 D+8) c_l^2 u_{l}+(D^2+6D+8) u_{l} \right)
\end{split}
\ee
where we have used $u_l = {\bm u} \cdot {\bm c}_l$, $u^2={\bm u} \cdot {\bm u}$,
 $c_l^2={\bm c}_l \cdot {\bm c}_l$.

\section*{Appendix B}

In this appendix we detail the technical issues of the boundary condition based on the combination of diffuse-reflection scattering kernel and the Newton-Raphson procedure \cite{NR}. We start from the kinetic boundary condition
\begin{equation}
f_l({\bm x}_W,t,u_w^{(eq)},T_w^{(eq)})
=\frac{
\sum_{l,{\bm c}_l\cdot{\bm n}\leq 0}
|{\bm c}_l\cdot{\bm n}|
f_l({\bm x}_W,t)}{
\sum_{l,{\bm c}_l\cdot{\bm n}>0}
|{\bm c}_l\cdot{\bm n}|
f_l^{(eq)}(u_w^{(eq)},T_w^{(eq)})}
f_l^{(eq)}(u_w^{(eq)},T_w^{(eq)}),
\ \ l,\ {\bm c}_l\cdot{\bm n}>0.
\end{equation}
For a  given time $t$, let us define two functions
\begin{equation}
\begin{split}
F_1({\bm x}_W,t,u_w,T_w;u_w^{(eq)},T_w^{(eq)})
=&
\rho u_w-\sum_{l}f_l c_l^x,
\\
F_2({\bm x}_W,t,u_w,T_w;u_w^{(eq)},T_w^{(eq)})
=&
2\rho T_w+\rho u_w^2-\sum_{l}f_l c_l^2.
\end{split}
\end{equation}
For prescribed $u_w$ and $T_w$, 
 we will find $u_w^{(eq)}$ and $T_w^{(eq)}$
 to satisfy $F_1=F_2=0$
 by means of the iterative Newton-Raphson procedure. The functions are expanded in Taylor series  with respect to generic variations $\delta u_w^{(eq)}$ and $\delta T_w^{(eq)}$ in the wall equilibrium velocity and temperature 
\begin{equation}
\begin{split}
&\left(\begin{array}{l}
F_1({\bm x}_W,t,u_w,T_w;u_w^{(eq)}+\delta u_w^{(eq)},T_w^{(eq)}+\delta T_w^{(eq)})\\
F_2({\bm x}_W,t,u_w,T_w;u_w^{(eq)}+\delta u_w^{(eq)},T_w^{(eq)}+\delta T_w^{(eq)})
\end{array}\right)
\\=&
\left(\begin{array}{l}
F_1({\bm x}_W,t,u_w,T_w;u_w^{(eq)},T_w^{(eq)})\\
F_2({\bm x}_W,t,u_w,T_w;u_w^{(eq)},T_w^{(eq)})
\end{array}\right)-
\frac{
\sum_{l,{\bm c}_l\cdot{\bm n}\leq 0}
|{\bm c}_l\cdot{\bm n}|
f_l({\bm x}_W,t)}{
\sum_{l,{\bm c}_l\cdot{\bm n}>0}
|{\bm c}_l\cdot{\bm n}|
f_l^{(eq)}(u_w^{(eq)},T_w^{(eq)})}
\left(\begin{array}{cc}
A&B\\
C&D
\end{array}\right)
\left(\begin{array}{r}
\delta u_w^{(eq)}\\
\delta T_w^{(eq)}
\end{array}\right)
\\&
+O((\delta u_w^{(eq)})^2)
+O((\delta T_w^{(eq)})^2)
+O((\delta u_w^{(eq)})(\delta T_w^{(eq)})),
\end{split}
\end{equation}
where the coefficients are
$$
A=\sum_{l, {\bm c }_l \cdot {\bm n} > 0} c_l^x\left( \frac{\partial f_l^{(eq)}}{\partial u_w^{(eq)}} \right)-\frac{\left( \sum_{l, {\bm c }_l \cdot {\bm n} > 0}|{\bm c}_l \cdot {\bm n}|\left( \frac{\partial f_l^{(eq)}}{\partial u_w^{(eq)}} \right)\right) \left(\sum_{l, {\bm c }_l \cdot {\bm n} > 0} c_l^xf_l^{(eq)} \right)}{\sum_{l, {\bm c }_l \cdot {\bm n} > 0} |{\bm c}_l \cdot {\bm n}| f_l^{(eq)}}
$$

$$
B=\sum_{l, {\bm c }_l \cdot {\bm n} > 0} c_l^x\left( \frac{\partial f_l^{(eq)}}{\partial T_w^{(eq)}} \right)-\frac{\left( \sum_{l, {\bm c }_l \cdot {\bm n} > 0}|{\bm c}_l \cdot {\bm n}|\left( \frac{\partial f_l^{(eq)}}{\partial T_w^{(eq)}} \right)\right) \left(\sum_{l, {\bm c }_l \cdot {\bm n} > 0} c_l^xf_l^{(eq)} \right)}{\sum_{l, {\bm c }_l \cdot {\bm n} > 0} |{\bm c}_l \cdot {\bm n}| f_l^{(eq)}}
$$

$$
C=\sum_{l, {\bm c }_l \cdot {\bm n} > 0} c_l^2 \left( \frac{\partial f_l^{(eq)}}{\partial u_w^{(eq)}} \right)-\frac{\left( \sum_{l, {\bm c }_l \cdot {\bm n} > 0}|{\bm c}_l \cdot {\bm n}|\left( \frac{\partial f_l^{(eq)}}{\partial u_w^{(eq)}} \right)\right) \left(\sum_{l, {\bm c }_l \cdot {\bm n} > 0} c_l^2 f_l^{(eq)} \right)}{\sum_{l, {\bm c }_l \cdot {\bm n} > 0} |{\bm c}_l \cdot {\bm n}| f_l^{(eq)}}
$$

$$
D=\sum_{l, {\bm c }_l \cdot {\bm n} > 0} c_l^2 \left( \frac{\partial f_l^{(eq)}}{\partial T_w^{(eq)}} \right)-\frac{\left( \sum_{l, {\bm c }_l \cdot {\bm n} > 0}|{\bm c}_l \cdot {\bm n}|\left( \frac{\partial f_l^{(eq)}}{\partial T_w^{(eq)}} \right)\right) \left(\sum_{l, {\bm c }_l \cdot {\bm n} > 0} c_l^2 f_l^{(eq)} \right)}{\sum_{l, {\bm c }_l \cdot {\bm n} > 0} |{\bm c}_l \cdot {\bm n}| f_l^{(eq)}}
$$
Neglecting the higher order terms and solving linear simultaneous equations with
$F_1({\bm x}_W,u_w,T_w;t,u_w^{(eq)}+\delta u_w^{(eq)},T_w^{(eq)}+\delta T_w^{(eq)})=
 F_2({\bm x}_W,u_w,T_w;t,u_w^{(eq)}+\delta u_w^{(eq)},T_w^{(eq)}+\delta T_w^{(eq)})=0$,
 we obtain the corrections
\begin{equation}
\begin{split}
&\left(\begin{array}{r}
\delta u_w^{(eq)}\\
\delta T_w^{(eq)}
\end{array}\right)
=
\frac{
\sum_{l,{\bm c}_l\cdot{\bm n}>0}
|{\bm c}_l\cdot{\bm n}|
f_l^{(eq)}(u_w^{(eq)},T_w^{(eq)})}{
(AD-BC)\sum_{l,{\bm c}_l\cdot{\bm n}\leq 0}
|{\bm c}_l\cdot{\bm n}|
f_l({\bm x}_W,t)}
\left(\begin{array}{rr}
D&-B\\
-C&A
\end{array}\right)
\left(\begin{array}{l}
F_1({\bm x}_W,t,u_w,T_w;u_w^{(eq)},T_w^{(eq)})\\
F_2({\bm x}_W,t,u_w,T_w;u_w^{(eq)},T_w^{(eq)})
\end{array}\right),
\end{split}
\end{equation}
which are added to the solutions 
\begin{equation}
\begin{split}
\left(u_w^{(eq)}\right)_{\rm new}=&
\left(u_w^{(eq)}\right)_{\rm old}+
\delta u_w^{(eq)},
\\
\left( T_w^{(eq)}\right)_{\rm new}=&
\left( T_w^{(eq)}\right)_{\rm old}+
\delta T_w^{(eq)}.
\end{split}
\end{equation}
The process is iterated to convergence.

\section{Acknowledgments}

We acknowledge useful conversations with R. Benzi, L. Biferale, A. Scagliarini  and S. Succi. M. Sbragaglia is also grateful to R. Surmas, C. E. Pico Ortiz and P.C. Philippi for useful suggestions received in an early stage of the writing.

\end{document}